\documentclass[epsf]{article}

\usepackage{amsmath,amsthm,amssymb}
\usepackage{graphicx}
\usepackage[footnotesize,nooneline,center]{caption2}
\usepackage{epsfig}
\usepackage[vflt]{floatflt}

\def \CC {{\mathbb{C}}}
\def \RR {{\mathbb{R}}}

\def \HH {{\mathbb{H}}}

\newtheorem{question}{\textnormal{\large{Q}\small{UESTION}}}[equation]

\newcommand{\keywords}[2]{\noindent\textbf{#1} : \emph{#2}.}

\begin{document}

\author{Alessandro Rosa}
\title{On a solution to display non-filled-in quaternionic Julia sets}

\maketitle

{Quaternionic Julia sets, iterates, dynamical
systems, graphics, escape time method, cut-off method,
illumination}

\vspace{0.5cm}

\begin{abstract}
During early 1980s, the so-called `escape time' method, developed
to display the Julia sets for complex dynamical systems, was
exported to quaternions in order to draw analogous pictures in
this wider numerical field. Despite of the fine results in the
complex plane, where all topological configurations of Julia sets
have been successfully displayed, the `escape time' method fails
to render properly the non-filled-in variety of quaternionic Julia
sets. So their digital visualisation remained an open problem for
several years. Both the solution for extending this old method to
non-filled-in quaternionic Julia sets and its implementation into
a program are explained here.
\end{abstract}

\section{Introduction}
The worldwide hype for holomorphic dynamics during the turn from
1970s to 1980s claimed the development of graphical methods to
display these systems on the computer screen, with emphasis to
complex Julia sets $\mathcal{J}$ \cite{VVAA-1988}. On the same
wave, thanks to the joint work by Mandelbrot and Norton, the first
pictures of quaternionic Julia sets came also out by applying the
same method as used in $\CC$ and known under the two definitions
of `\emph{escape time}' or `\emph{level sets}'. Few works, either
in graphics terms, were published on this subject since early
1980s: namely, the book \cite{Mandelbrot-1983} and the articles by
Norton \cite{Norton-1982,Norton-1989}, Holbrook
\cite{Holbrook-1983,Holbrook-1987}, by Hart, Kauffman and Sandin
\cite{HSK-1989,HSK-1990}. For analogous studies in higher
dimensional dynamics, see \cite{DiStBu-1996}. Quaternions are
multi-dimensional values consisting of 4 vectors in the form
\begin{equation}\label{IntroEq1}
h=r+mi+nj+pk,\qquad r,m,n,p\in\RR,\qquad i^2=j^2=k^2=-1
\end{equation}
and generate a non-abelian group. The quaternion space is named
$\HH$ for convention, in honor of William R. Hamilton
(1805--1865), inventor of such numerical values for solving some
issues arising from rotation in 3-D Euclidean spaces.

\begin{figure}[htb]
\vspace{0.3cm} \centering
  \input{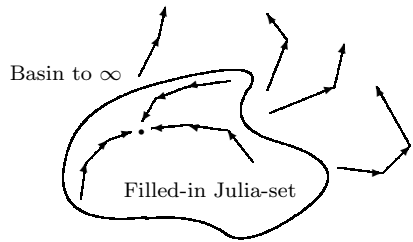}
  \setcaptionwidth{0.9\textwidth}
  \captionstyle{normal}
  \caption{\footnotesize \textbf{Filled-in Julia sets $J$.} Both
  topological and dynamical were explained after the first explorations of iterates in the
  complex numbers, by means of the quadratic map $f(z):z^2+c$. Orbits outside $J$
  converge to $\infty$ and they are best viewed in the Riemann sphere model of picture \ref{qhd23}.}
  \label{qhd07}
\end{figure}
The `\emph{escape time}' method enjoys one major limitation: as
one tries to display, for example, the \emph{non-filled-in} Julia
sets of basins $\mathcal{B}$ of attraction to a finite limit fixed
point $\delta$ and extending to $\infty$ (often retrieved by the
iterates of rational maps). This article explains how to modify
the `standard'\footnote{We liked to use this naming convention in
order to differ the `\emph{escape time}' method from our new
version.} method so to fill this leak and extend it to the
`non-filled-in' case. After resuming the technical details of the
standard method in section \ref{Background}, we discuss the
novelties in section \ref{Considerations}; finally, the
implementation into the program QHD is explained in section
\ref{QHD}.

\section{The background}\label{Background}
\subsection{Julia sets topology}
Up to complex numbers and given a rational map
$$f=\frac{P}{Q}$$
where $P,Q$ are co-prime polynomials, i.e. with no common roots,
it was observed that Julia sets $J$ (in any number field) may:

\begin{enumerate}
    \item be (pointwisely) totally disconnected;
    \item have $n-1$ dimensions;
    \item have $n$ dimensions.
\end{enumerate}

\begin{question}
Could this classification hold for any $n$-dimensional number
field?
\end{question}

In $\RR$, where $n=1$, points 1) and 2) are the same, so that real
Julia sets $\mathcal{R}$ are pointwisely totally disconnected (in
fact, for point 2), we have: $n-1=0$) or the whole real line (for
point 3), $\mathcal{R}$ is $n=1$ dimensional).

In $\CC$, where $n=2$, $\mathcal{J}$ could be either a dust of
points (0-dimensional), also a line ($n-1=1$ dimensional) or an
area ($n=2$ dimensional).

As far as the author knows, it is not known whether point 1) holds
for quaternionic $\mathcal{J}$ too, or they are just disconnected
sets with (hyper-)volumes. We will implicitly assume here the
first two cases only, discarding \emph{a priori} the third one,
which leads to the trivial case where $\mathcal{J}$ spans over the
whole $4$-dimensional space.

\subsection{The open problem}
For complex numbers, any Julia set $\mathcal{J}$ can be
successfully displayed with the already available methods; new and
customised methods are anyway required to explore local vector
fields\footnote{Like for the so-called `\emph{hedgehogs}',
invariant sets about irrationally indifferent fixed points not
enjoying the diophantine condition or rational approximation; the
digital display was also an open problem here. A related graphical
method arranging a good solution is going to be explained in
\cite{ChRo-2006}.} about the fixed points $\delta$. But
quaternionic $\mathcal{J}$ did not share the same luck. The method
employed in $\HH$ was radically borrowed from $\CC$. However, as
the related literature of those times indicates -- together with
the book \cite{DaKa-1997-1}, the road-map was to display the
iterations of polynomials and to develop the details enhancement:
in fact, those graphical examples always concerned of disconnected
or of the filled-in varieties $\mathcal{J}$ only, mostly coming
from the family of general quadratic polynomials
\begin{equation}\label{OpenEq1}
ph^2+q,\qquad\qquad h,p,q\in\HH.
\end{equation}

\begin{figure}[htb]
\centering \epsfxsize=3.4cm
  \epsffile{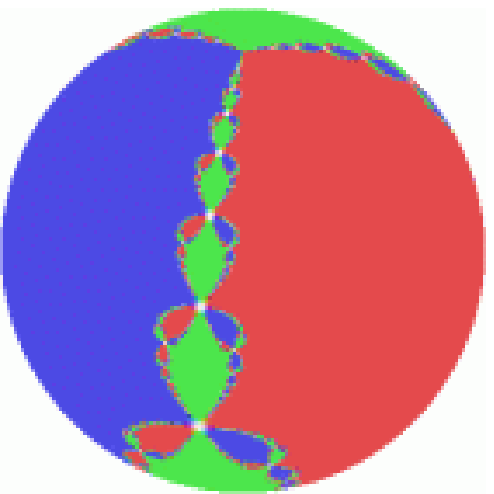}
  \setcaptionwidth{0.6\textwidth}
  \captionstyle{normal}\caption{\textbf{Non-filled-in example.} A case arising for the map (\ref{NewtonEq}): this view is taken on the Riemann sphere, in the neighborhood of the point at infinity.}\label{qhd23}
\end{figure}

Julia sets are mostly a mathematical question. This flaw in
graphical terms was just the side effect from the interests of
those pioneering researchers, who mainly wanted to start from the
easier properties of polynomials\footnote{In the course of these
early mathematical explorations, they proved that quaternionic
$\mathcal{J}$ display in 3 dimensions as rotations of the related
$\mathcal{J}$ in $\CC$ around the X-axis.}. Quaternions are
non-commutative and the study of rational maps in $\HH$ may get to
high level of complication, in mathematical terms: so the
investigation on these maps was not pursued. So it came that the
graphics development was not considered along that direction:
there was no need to quest on the central question of this article
throughout the older works, because all topological cases of Julia
sets for polynomials were already fulfilled by the standard
method. (The complete related literature was listed in the
bibliography.) Thus we can reasonably state that this problem was
still open. A \emph{filled-in} Julia set $J$ includes all seed
points whose forward iterated orbits do not tend\footnote{This is
not the same as stating that these orbits should converge to a
constant limit fixed point $\delta$, due to the complicate
dynamics -- where one cannot properly speak of convergence -- in
the neighborhood of indifferent fixed points, both for the
rational and irrational case.} to $\infty$. Topologically
speaking, the boundary of a filled-in $J$ is homeomorphic to a
closed and bounded curve (even with fixed points). But exceptions
are thrown when $J$ also extends to $\infty$: for example, in the
classic example of the cubic rational map, resulting from Newton's
method applied to the complex function $f(z):\ z^3-1$,
\begin{equation}\label{NewtonEq}
f(z):\frac{2z^3+1}{3z^2}.
\end{equation}
This variety of Julia sets $J$ will be said
\emph{`non-filled-in'}: this case is out of the range of
possibilities for the `escape time' method when applied to
quaternions.

Being $\mathcal{J}$ displayed as a three dimensional model, the
latest improvements, dating back to early 1990s (see references
list), mainly concerned of the application of efficient
illumination techniques or of improving the boundary refinement.
Successful results in the former direction were achieved by means
of Phong illumination model \cite{FDFH}; whereas the reader may
consult \cite{HSK-1989,HSK-1990} about the second issue. Today
software performs higher quality pictures of $\mathcal{J}$ than
ever before, but these programs\footnote{In fig. \ref{TableGintz},
the reader can watch the earliest results from the application of
our method by one of such high quality programs.} still rely on
the standard method: thus only the filled-in $\mathcal{J}$ can be
properly displayed. After deeply scanning the Web and
consultations with experts in this topic, programs displaying the
non-filled-in family have been not obviously coded yet. Author's
solution wants to overcome this limitation; the related code has
been already implemented and tested to work into one PC program
`QHD', freely downloadable from author's
site\footnote{http://malilla.supereva.it} and applied to render
some among the further pictures.

\subsection{The standard method}
From (\ref{IntroEq1}), one understands that Julia sets
$\mathcal{J}$ exist in four dimensions. So one is not able to
figure them in mind or to display somewhere. One is just able to
embed them into two ($\RR^2$) or three ($\RR^3$) real Euclidean
spaces onto the computer screen. In this sense, different
two-staged methods are available but in general they all root to a
first stage where values/points are computed and to a second one,
when the screen points are colored. In $\CC$ colors palettes may
be arbitrarily set (even to render purely aesthetic effects), but
in $\HH$ rigor shall be followed because $\mathcal{J}$ shall give
their solid model look.

\begin{figure}[htb]
\centering
  \epsfxsize=5.0cm
  \epsffile{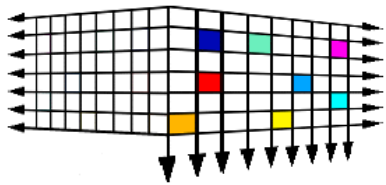}
\setcaptionwidth{0.7\textwidth}
  \captionstyle{normal}\caption{A region of the 3-D Euclidean space is scanned, so that each point is then
  associated to one quaternion (by setting the X,Y,Z values to $r,m,n$ respectively and $p=0$).}\label{qhd24}
\end{figure}

\noindent In practice, this translates into this three steps
approach:

\begin{enumerate}
    \item to \emph{scan} a given region, provided a discretization of it
in order to have a $n-$dimensional grid of values $p$;
    \item to \emph{iterate} each $p$ for a number of times $k$;
    \item to \emph{test} the resulting iterated value $p_k$ in order to
color them differently or choose which ones have to be plot.
\end{enumerate}

\begin{figure}[htb]
\centering
\begin{tabular}{ccc}
  \epsfxsize=2.5cm\epsffile{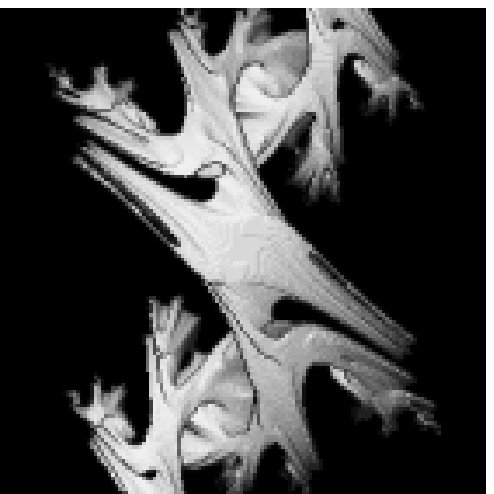} &
  \epsfxsize=2.5cm\epsffile{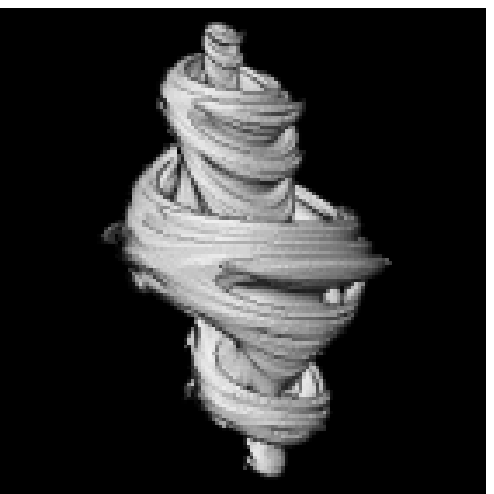} &
  \epsfxsize=2.5cm\epsffile{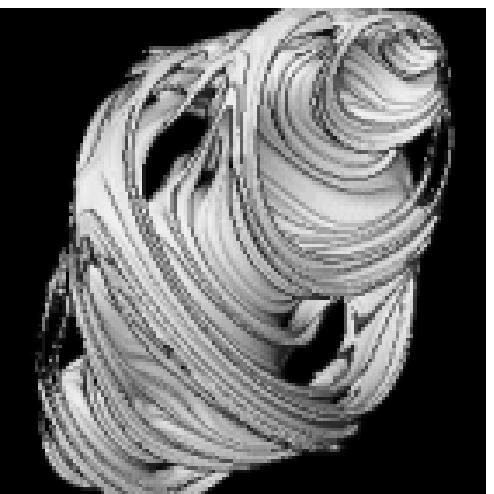}\\
\end{tabular}
  \caption{Examples of filled-in Julia sets $\mathcal{J}$ for the quadratic $h^2+c$.}\label{qhd10-12}
\end{figure}

The `standard' method was applied by Mandelbrot and Norton and by
Holbrook in their first visualizations and displays the
3-dimensional models of $\mathcal{J}$. We will follow the same
line here, focusing on Julia sets $\mathcal{J}$ possibly showing
up as totally disconnected or 3-D surfaces. The care of
illumination is fundamental for any 3-D model, in order to prevent
the flat looking. Phong approach guarantees high performance, for
example, with regard to details, enhanced as the shining effect is
rendered very accurately and very close to real light. In general
the standard method, applied to $\mathcal{J}$, is son of what
usually denoted `\emph{escape method}' in the early literature
\cite{VVAA-1988} devoted to the digital visualization of
$\mathcal{J}$. We pointed out that malfunctions come up for
non-filled-in $\mathcal{J}$ from iterates of rational maps, for
example from the application of the Newton-Raphson's method. In
this case, one might find out those basins of attraction to a
finite limit point and extending to $\infty$. See p.
\pageref{NewtonEq} including a picture from the cubic quaternionic
function $f(h):h^3-1$, which turns into:
\begin{equation}\label{NewtonEq1}
    T_f=h-\frac{f(h)}{f'(h)}=\frac{2h^3+1}{3h^2}.
\end{equation}
The Julia set curve $\mathcal{J}$ includes infinitely many fixed
points $\delta$ and also extends to infinity. As the standard
method applies to (\ref{NewtonEq1}), one sees fig. (\ref{qhd25}).
One also notices how roughly it works in this situation: the shape
of $\mathcal{J}$ is not so clear and just a part of it is
displayed, but imprecisely. The grey region is the evidence of the
failure of the standard method, which cannot adequately check the
sequences of iterated values here.

\subsection{Tuning the parameters}
Before entering the heart chapter, one should be assured that a
modification really urges. So we tried to tune the two parameters
involved in this kind of experiments: the bailout disc radius and
the number of iterations. Finally, we reported the resulting
pictures in table \ref{TableTune}. This experiment is relevant
because, for sake of completeness, one shall either \emph{show the
reasons why something is required and why it is not}: these
pictures want to be the empirical evidence of what will be further
discussed in more details at section
\ref{starting_considerations}.
\begin{table}
  \centering
    \begin{tabular}{ccccccc}
    \footnotesize\em $r = 0.8$ &\hspace{-1.6cm} & \footnotesize\em $r = 1.0$ & \hspace{-1.6cm} & \footnotesize\em $r = 2.0$ & \hspace{-1.0cm} & \footnotesize\em $r = 4.0$\\
    \epsfxsize=2.3cm\epsffile{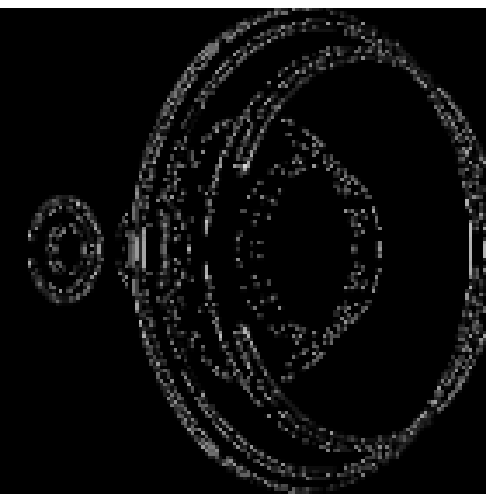} & &\epsfxsize=2.3cm\epsffile{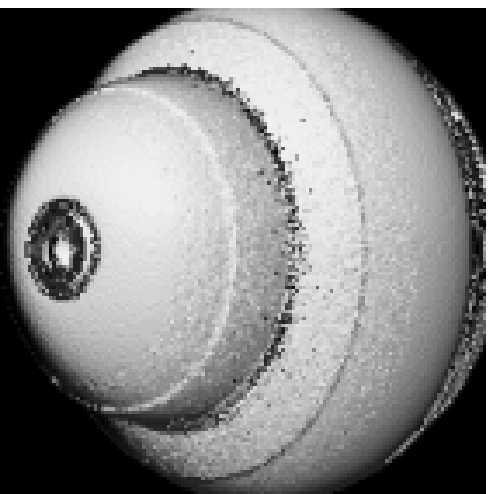} & & \fbox{\epsfxsize=2.3cm\epsffile{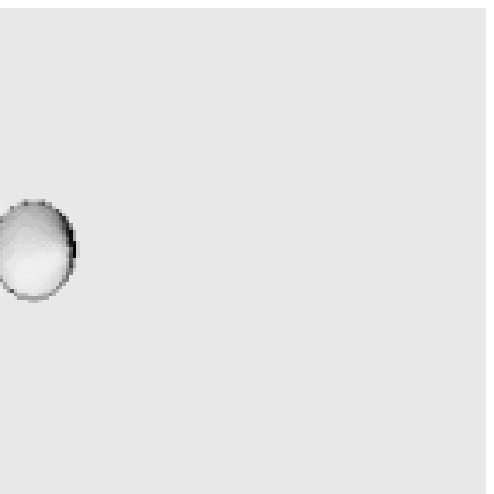}} & & \fbox{\epsfxsize=2.3cm\epsffile{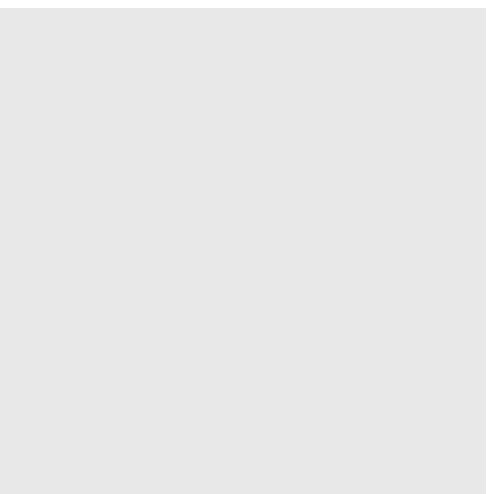}}\\
    \epsfxsize=2.3cm\epsffile{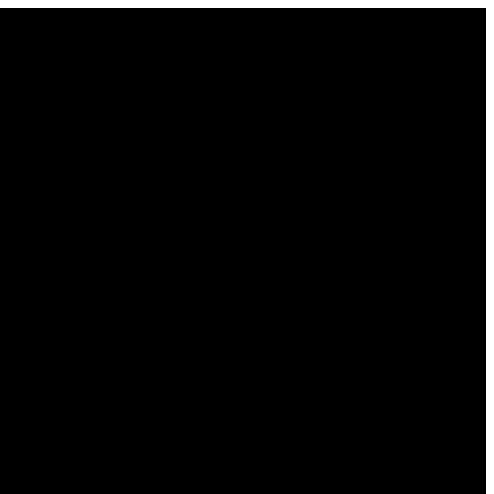} & &\epsfxsize=2.3cm\epsffile{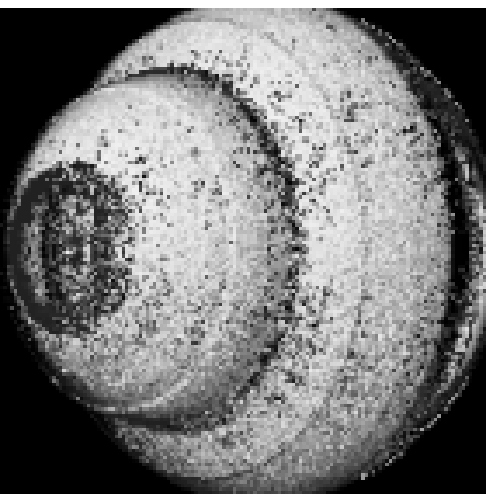} & & \fbox{\epsfxsize=2.3cm\epsffile{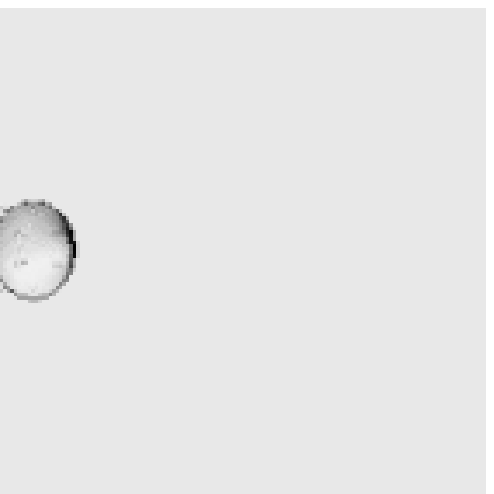}} & & \fbox{\epsfxsize=2.3cm\epsffile{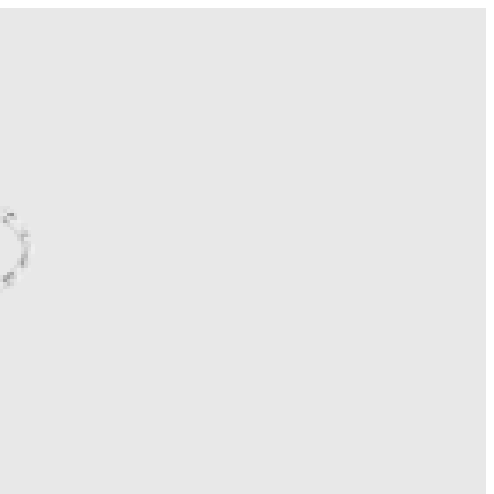}}\\
    \epsfxsize=2.3cm\epsffile{qhd20a} & &\epsfxsize=2.3cm\epsffile{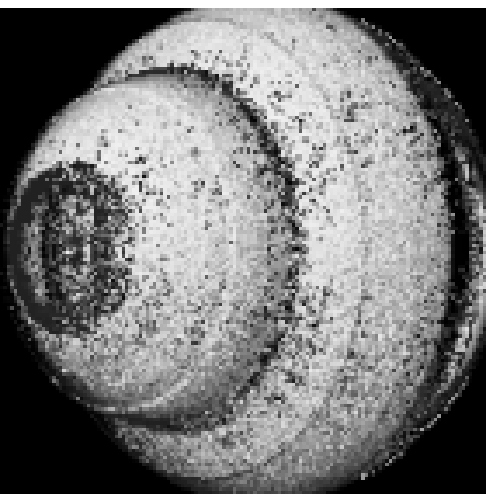} & & \fbox{\epsfxsize=2.3cm\epsffile{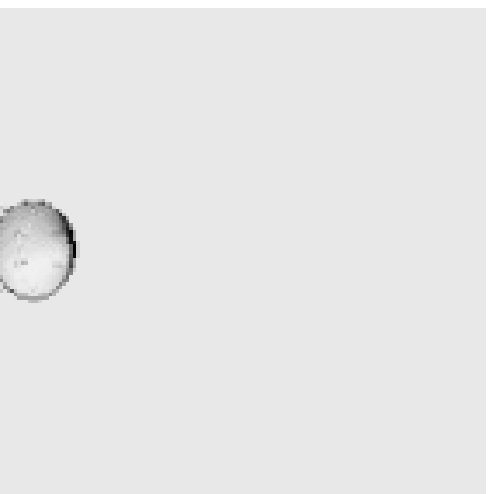}} & & \fbox{\epsfxsize=2.3cm\epsffile{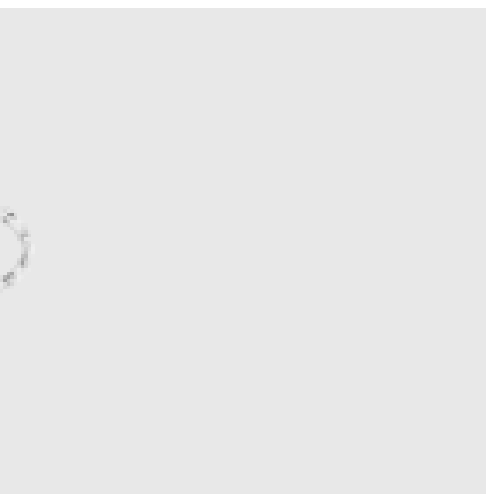}}\\
    \multicolumn{7}{c}{\fbox{\epsfxsize=2.5cm\epsffile{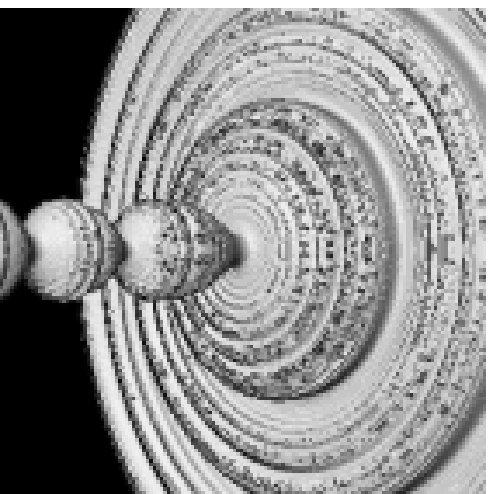}}}\\
    \end{tabular}
    \setcaptionwidth{0.8\textwidth}\captionstyle{normal}
  \caption{\textbf{Experiments with the standard method.} All top pictures have been rendered by `escape time' method.
  Columns relate to bailout radius, whereas rows to iterates 5, 12, 24. The bottom picture was computed by the new `cut-off rate' approach
  for watching the correct rendering.}\label{TableTune}
\end{table}
The table \ref{TableTune} at p. \pageref{TableTune} includes
figures of the same Julia set $\mathcal{J}$, the large figure at
the bottom is correctly drawn by the `cut-off rate' approach and
shall be compared with the small ones, rendered by the `escape
time': the number of iterates and the bailout circle radius
(centered at the origin in the classical method) were tuned to
distinct values. In some cases, one finds out a solid piece (the
radius at 1.0 was even too large and gave a great approximated
shape); for larger radii, the grey region indicates that the
location of $J$ cannot be distinguished from the basin locus,
because the detection rule failed\footnote{See section
\ref{CrashCourseReasons} for technical details.}. Finally, the
black pictures denote that the chosen conditions were rather
insufficient.

\section[Opening quaternionic Julia sets to rational maps]{Opening quaternionic Julia sets\\to rational maps}\label{Considerations}
\subsection{Disclaiming the Analysis}
After implementing the currently available methods to display
quaternionic Julia sets $J$ for polynomials, the author wanted to
achieve the same even for iterated rational maps. In mathematical
terms, the study of the algebraic properties related to them may
be harder than in $\RR$ or in $\CC$, owing to the
non-commutativity property enjoyed by quaternion numbers. The next
sections have been deliberately compiled to solve a practical
problem exclusively, not regarding the mathematics behind the
nature of the functions involved. In fact, the question on
quaternionic Julia sets connectivity still deserves a theoretical
systematisation, because the arising topological configurations
are not quite the same as in $\CC$, for example disconnected Julia
sets in $\HH$ do show connected subregions, whereas they are
totally disconnected (Cantor dust) in $\CC$. This also invalidates
the same definition of Mandelbrot set as it holds in $\CC$. But
this goal is out of scope for this article and it goes beyond
author's knowledge.
\begin{figure}[htb]
  \begin{minipage}[t]{0.5\textwidth}
  \centering
  \fbox{\epsfxsize=3.2cm\epsffile{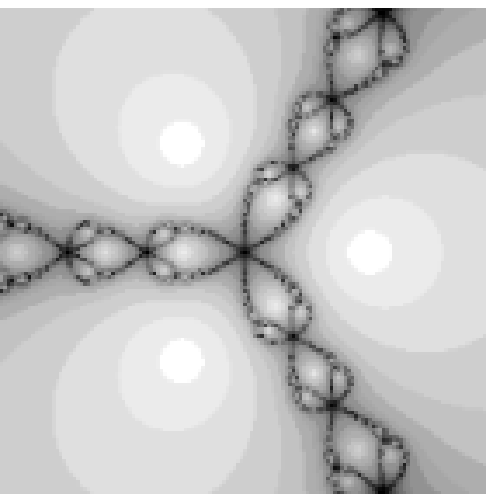}}
  \caption{Julia set for (\ref{NewtonEq1}) in $\CC$.}\label{qhd30}
  \end{minipage}
  \begin{minipage}[t]{0.5\textwidth}
  \centering
  \fbox{\epsfxsize=3.2cm\epsffile{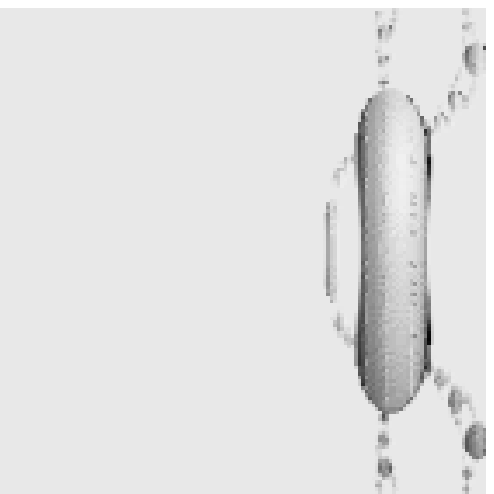}}
\setcaptionwidth{0.9\textwidth} \captionstyle{normal}
  \caption{\textbf{Malfunctions.} The quaternionic Julia set generated by (\ref{NewtonEq1}) and
  processed by the standard method.}\label{qhd25}
  \end{minipage}
\end{figure}

\subsection{Starting considerations}\label{starting_considerations}
The standard method scans the given area (in $\CC$) or volume (in
$\RR^3$, the reduced vector space from $\HH$), associating a
numerical value per each element of this region; the value is
iterated for a number of times under a given function. If the
resulting value is still inside a sphere of given radius, then the
orbit is said to be `trapped' and then it belongs to a basin of
attraction bounded by a Julia set $\mathcal{J}$. But this only
works well when $\mathcal{J}$ is a closed curve which the test
disc is able to catch completely up; on the contrary, the disc
does not fit any more for the whole $\mathcal{J}$ extending to
$\infty$. Historically speaking, the standard method inherits the
obvious limitation from the approach devoted to display Julia sets
$\mathcal{J}$ on a computer and whose location and topology inside
$\CC$ were \emph{a priori} known, either if filled-in or
disconnected. In fact, the iterated maps, whose $\mathcal{J}$ were
going to be shown onto a computer screen for the first times ever
during early 1980s, were quadratic polynomials and namely in the
form
\begin{equation}\label{Eq6}
z^2+c,\qquad z,c\in\CC;
\end{equation}
here all cases of $\mathcal{J}$, their topology and the attracting
fixed points are known to approximately lie inside a bailout disc
with radius $2$ at most. The standard method was invented with
these `bounded' dynamics in mind and it fits the range of cases
offered by (\ref{Eq6}) only; thus, in this sense, it is a
customised method which unfits the cases not covered by its
possibilities.

The standard method can be reset by changing the radius of the
trapping disc in $\CC$ but limitations come up as one deals with
quaternions: such method strictly depends on the same
functionality of the bounded ball (as the concept of trapping disc
turns into for the embedding of $\HH$ into $\RR^3$): i.e. when
orbits are tested to stay bounded or to escape to $\infty$; in
fact the same method implicitly assumes that there is only
\emph{one} finite and attracting fixed point and that its
location, together with the related basin, is \emph{a priori}
known to lie inside the test ball itself. So, for more general
purposes -- i.e. \emph{regardless of fixed points locations and
quantity} -- the standard method does not work well.

This will be the starting point upon which the enhancement will be
based.

\begin{figure}[htb]
  \begin{minipage}[t]{0.45\textwidth}
  \centering
  \input{qhd17.pic}
  \captionstyle{normal}\caption{\textbf{First step.} Quaternionic Julia set boundary detection.}\label{qhd17}
  \end{minipage}
  \begin{minipage}[t]{0.5\textwidth}
  \centering
  \input{qhd18.pic}
\setcaptionwidth{0.85\textwidth} \captionstyle{normal}
  \caption{\textbf{Second step.} The recurrent process for approximating the Julia set point location.}\label{qhd18}
  \end{minipage}
\end{figure}

\subsection{Crash course reasons}\label{CrashCourseReasons}
Therefore why does the standard approach fail here?

First we say why it does not in $\CC$: actually because all points
of a given region are iterated and tested, regardless of their
location and even visibility is not a relevant condition; in fact
one draws both the basin and the Julia set in $\CC$. But in 3-D,
one needs to differ them because the basins shall be not drawn or
one would just watch a cubic block instead ! The points of
quaternionic $\mathcal{J}$ do need to be scanned and sorted if
they are visible to be plot or not. This is the main direction
which methods for quaternionic Julia sets must follow straight.
The exclusion of the basins during the plot is a delicate feature.

`Filled-in' or disconnected Julia sets configurations keep an
element making the difference: the basin $\mathcal{B}_\infty$ to
infinity. Its role is clearer in the `filled-in' case (fig.
\ref{qhd07}/B): one plays a boolean situation where points of
$\mathcal{B}_\infty$ exit the test sphere, others do not. The rule
is that $J$ is met when, given two close seeds to $J$, the related
orbits are checked by the `escape time' and one finds that an
orbit runs to infinity, the other to the finite attracting point
(fig. \ref{qhd17}). So the spatial interval, wherein a point of
$\mathcal{J}$ should be approximately located, shrinks over and
over again up to lower bound, while the fates of the two orbits
keep different (fig. \ref{qhd18}).

\begin{figure}[ht]
\centering
  \input{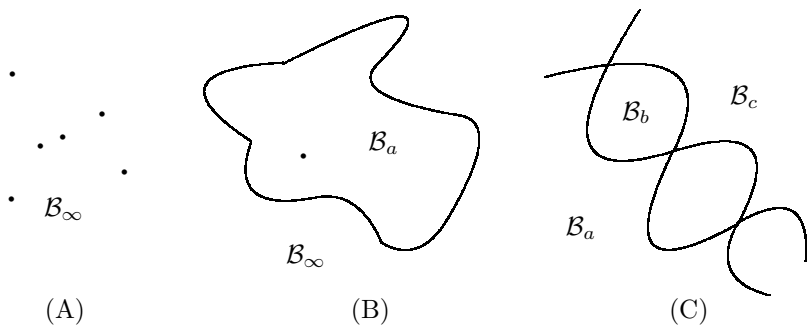}
  \setcaptionwidth{0.85\textwidth}
  \captionstyle{normal}
  \caption{\footnotesize The above figures resume three topological
  configurations for complex Julia sets: (A) totally disconnected and
  with the only basin to infinity, (B) connected and
  splitting $\CC$ into two basins only, a basin to infinity and one to the finite
  fixed point. Then (C) a Julia set and all basins (two or more) of attraction to
  finite points.}
  \vspace{0.3cm}
  \label{qhd16pic}
\end{figure}
The standard method crashes because, for such maps as
(\ref{NewtonEq1}), infinity is no longer attracting and
$\mathcal{J}$ splits $\HH$ into basins of attraction to finite
$\delta$: as $\mathcal{J}$ is even crossed, it is not detected
because neighboring orbits never escape the test sphere, and the
above rule blows away (see fig. \ref{qhd25}).

\subsection{The `cut-off rate' method}
The modification mainly relies upon the same region scanning
process like in the standard method, but it adds the previous
considerations, together with the inspiration from the application
of normal families to holomorphic dynamics.

Let $f$ be an holomorphic function, then
$$f^0(z)=z, f^1=f(z), \dots f^n(z)=f^{n-1}(f(z))$$
are said to be the iterates of rank $0, 1, \dots, n$. From
iteration theory, one knows that, for a same attracting fixed
point $\gamma_k$, the total basin $\mathcal{B}_{\gamma_k}$ of
attraction is the union set of all the points whose orbits
converge to $\gamma_k$:
$$\lim_{n\rightarrow\infty}f^n(z)=\gamma,\qquad \forall z\in\mathcal{B}_{\gamma_k}.$$

\begin{figure}[htb]
      \centering
      \input{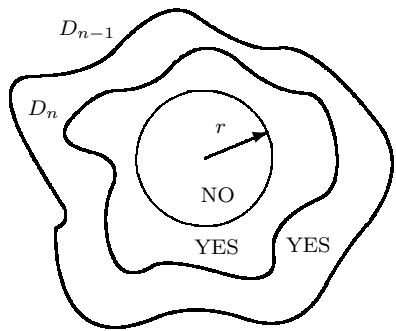}
      \vspace{0.3cm}
      \setcaptionwidth{0.7\textwidth}
      \captionstyle{normal}
      \caption{Iterates whose distance is smaller
      than a given value are not considered for the final plot.}\label{qhd26}
\end{figure}

Let $\mathcal{D}_1, \mathcal{D}_2, \dots \mathcal{D}_n,$ be Jordan
curves bounding simply connected domains $D_1, D_2,$ $\dots, D_n$
around $\gamma_k$ respectively (see fig. \ref{qhd26}), so that
$D_{n-1} \supset D_n$; thus, in metric terms, the distance $d_n$,
of any pair of points $p_{n-1}\in D_{n-1}, p_n\in D_n$ (points of
an orbit), shrinks to $0$ as the orbit itself gets closer to
$\gamma_k$:
\begin{equation}\label{Eq2}
d_n=|p_n-p_{n-1}|<\epsilon,\qquad \epsilon > 0, \lim\epsilon=0.
\end{equation}

So, generally speaking, given a seed point
$w\in\mathcal{B}_{\gamma_k}$ and close to $\mathcal{J}$, the
distance $d_n\rightarrow 0$ as the iterative index
$n\rightarrow\infty$. Conversely -- very important to understand
how the modification works --, one may also state that if $d^n$ is
set to a constant $s$, then one can re-elaborate (\ref{Eq2}) into
$$|f_n(w) - f_{n+1}(w)|>s$$
as the seed $w$ gets closer and closer to $\mathcal{J}$. This
offers the chance to part all possible orbits generated throughout
one basin $\mathcal{B}_{\gamma_k}$ into two groups:

\begin{enumerate}
    \item the orbits of the seed points $p$ which converge, to the fixed
point $\gamma_k$, under a given distance $d_n$ and after a smaller
number of iterations;
    \item other orbits of seeds $p$ which need a larger number of
iterates to converge under $d_n$.
\end{enumerate}

The points being closer to $\mathcal{J}$ belong to the second
group. Hence the goal of revisiting the standard method is the
`isolation' of the second group of points and then plot them
exclusively, with regard to other basins distributions where even
the test ball of the standard method, with fixed location, does
not fit any longer to return clear and good graphical results. If
the test ball condition is:
\begin{equation}\label{Eq3}
    |p_n-p_{n-1}|<r.
\end{equation}

This isolation can be simply performed by assuming the reverse
inequality of (\ref{Eq2}), that is:
\begin{equation}\label{Eq4}
|p_n-p_{n-1}|\geq r.
\end{equation}

The operator `$\geq$' achieves a reverse task: rather than
trapping, external orbits to the ball are assumed or in other
terms, the successive iterated domains $D_n$ exclusively, whose
distances $d_n\geq r$. That is, the domains of points which are
farther, in terms of iterates, from $\gamma_k$: conversely
speaking, the points which are closer and closer to Julia sets.
Moreover one drops off the test condition of the escape method,
about the upper bound of the iterative index, which prevent
infinite loops; we preferred to adopt the first member of
inequality (\ref{Eq4}), because it showed to work more finely: the
major enhancement is to offer, by means of just one condition of
convergence, the work on $n$ test balls, as many as the number of
attracting fixed point $\gamma_k$ of the iterated map. Therefore
the reader has no longer to deal with the issue of knowing a
priori the number and the location of the attracting fixed points
$\delta$ or how long one orbit takes to converge in some
sufficiently close neighborhood of the fixed point. One sees such
test balls, \emph{which do not depend on the location anymore},
rising up from (\ref{Eq4}) and they work at the same time without
their centers to be preset \emph{ad hoc} in the algorithm code.
This condition grants any arbitrary input map.

The black regions in fig. (\ref{qhd27}) indicate that those points
have not been deliberately plot, as result from the isolation task
achieved by the `cut-off rate' method.
\begin{figure}[ht]
  \centering
  \fbox{\epsfxsize=2.8cm\epsffile{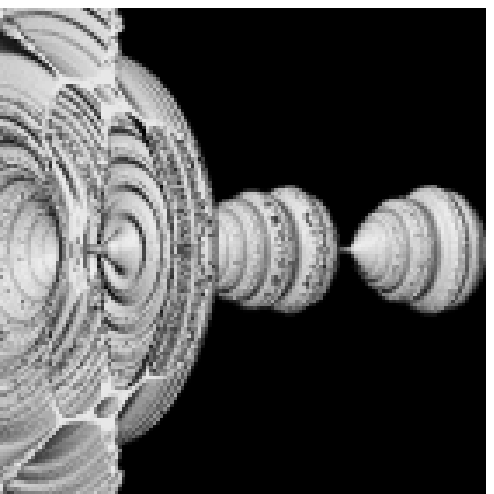}}
  \setcaptionwidth{0.8\textwidth}
  \captionstyle{normal} \caption{\textbf{Cut-off rate method.} The quaternionic Julia set for $$\frac{2h^3-2h^2-3}{3h^2+4h+1}.$$
  The set was sectioned to notice either the rotational symmetry with the Julia set in $\CC$ and
  the existence of interweaving basins which would make the classic method crash.}\label{qhd16}
\end{figure}

\section{Related software}\label{RelatedSoftware}
In this section we will give a short overview on some related
software which is currently implementing the `cut-off rate'
method.

\subsection{QHD : Quaternionic Holomorphic Dynamics}\label{QHD}
First we show one author's program, coded to make the early
experiments on the `cut-off rate' method: QHD is a branch of
another application, coded years before and named `Inwards to
Chaos', which is devoted to iterations in real, complex and
quaternion fields. In order to focus on the visualisation
techniques, QHD came up and became the natural environment to
display the graphical results related to the `cut-off rate'
method.

\subsection{The visual interface}\label{VisualInterface}
This application opens with a user-friendly interface to choose
the set to draw (`Julia' or `Mandelbrot'), the drawing algorithm
(`Standard' or `Extended'), a default set of formulas and the
illumination method (`\emph{Simple Lambertian}',
`\emph{Lambertian}' and `\emph{Phong}').

When properties are displayed, the same window expands and shows
additional parameters. First the details level yielding as finer
pictures as it is incremented (as well as computation times). Then
the number of iterates, the degree of predefined maps, the region
coordinates and the value of parameter $q$, like in formula
\ref{OpenEq1}).

Input features also list the command to load/save the parameters
and the metrics related to Julia set configuration, such as the
bailout value playing as the radius of the disc entrapping the
orbits in the standard method or as the value $r$ in the
inequality (\ref{Eq4}) of the extended approach.

The main interface offers a very little preview shot by a quick
plot, but one can display magnifications into a larger window; a
set of four arrows allows finally to change the input region
coordinates.

\begin{figure}
  \centering
  \begin{tabular}{c}
  \fbox{\epsfxsize=2.5cm\epsffile{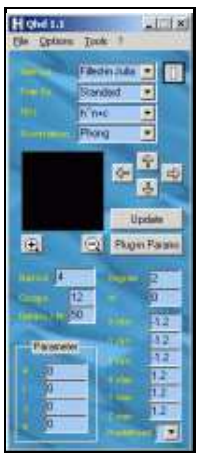}}\label{qhd01}
  \end{tabular}
  \caption{On the left, the main interface of QHD.}\label{TableSubmersions}
\end{figure}

Additional features include a player to stop the drawing process
manually, a window to rotate the given object along the three
X-Y-Z axes, together with one panel to manage all parameters
related to the illumination models (location of light points,
intensity, ambient diffusion). Then, there is the possibility of
setting any triplet of the four quaternionic components into the
X-Y-Z coordinates system in order to watch the possible shapes
assumed by quaternionic Julia sets into a reduced 3-D vector space
(see examples in table \ref{TableSubmersions} at p.
\pageref{TableSubmersions}). The output allows to save the current
picture into one graphics format file or to be copied into the
computer clipboard for exchanging it with other applications.

\vspace{0.3cm}

\subsection{Looking around}
My goal was to test if that method really worked in other
applications outside mine and verify if my results were achieved
in particular situations or due to possible mistakes in the code.
In the recent period (February 2006), after testing the method on
QHD, I wanted to let the `cut-off' method circulate among groups
interested in computer graphics devoted to quaternionic Julia
sets, which are usually frequented by `mathematized' programmers,
that is, non-mathematicians, strictly meaning, but skillful people
liking to deal with Mathematics. One is Terry W. Gintz who
welcomed the new method and implemented it into a program,
retrieving high-quality renderings of such fractals; for example
see the pictures in the table \ref{TableGintz} at p.
\pageref{TableGintz}.

\section{Conclusions}\label{conclusions}
The program `QHD' can be freely downloaded from the web-site
address at \texttt{http://www.malilla.supereva.it}; Opinions and
suggestions are welcomed to enhance it. The C++ code can be asked
via author's e-mail.

\vspace{0.5cm}

\begin{tabular}{p{6.0cm}l}
& \small Alessandro Rosa\\
& \small Brindisi, Italy\\
& \small \texttt{zandor$\_$zz@yahoo.it}\\
\end{tabular}

\bibliographystyle{amsplain}

\end{document}